\documentclass[aps,prd,twocolumn,showpacs,groupedaddress]{revtex4}
\usepackage{epsfig}
\usepackage{longtable}
\usepackage{amsfonts}
\newcommand{\Det}{\mbox{det}}

\newcommand{\oo}{{\omega_\oplus}}

\newcommand{\To}{T_\oplus}

\newcommand{\oxi}{\overline \xi}
\newcommand{\tr}{{\rm tr}}
\newcommand{\be}{\begin{equation}}
\newcommand{\ee}{\end{equation}}
\begin{document}
%\draft
%\preprint{HEP/123-qed}

\title{Testing Lorentz invariance by use of vacuum and matter filled cavity resonators}
\author{Holger M\"{u}ller}
\affiliation{Physics Department, Varian Bldg., Room 226, Stanford University, Stanford, CA 94305-4060; Phone: (650) 725-2354, Fax: (650) 723-9173} \email{holgerm@stanford.edu}

\date{\today}

\begin{abstract}
We consider tests of Lorentz invariance for the photon and fermion sector that use vacuum and matter-filled cavities. Assumptions on the wave-function of the electrons in crystals are eliminated from the underlying theory and accurate sensitivity coefficients (including some exceptionally large ones) are calculated for various materials. We derive the Lorentz-violating shift in the index of refraction $n$, which leads to additional sensitivity for matter-filled cavities ; and to birefringence in initially isotropic media. Using published experimental data, we obtain improved bounds on Lorentz violation for photons and electrons at levels of $10^{-15}$ and below. We discuss implications for future experiments and propose a new Michelson-Morley type experiment based on birefringence in matter.
\end{abstract}

%\tableofcontents

\pacs{11.30.Cp, 03.30.+p, 06.20.-f, 04.80.Cc}
%\keywords{}

\maketitle

%\tableofcontents
\section{Introduction}
 
Small violations of Lorentz invariance are allowed in a broad range of theories aiming for a unification of quantum mechanics and gravity and are one of the few possible signatures of quantum gravity in laboratory experiments. A description of such Lorentz violation is given by the Standard Model Extension (SME) \cite{ColladayKostelecky}: Based on a Lagrangian formulation of the Standard Model that incorporates the known particles and interactions, it adds all terms that can be formed from the particle fields and Lorentz tensors. In these additional terms, violation of Lorentz (rotation and boost) invariance is encoded. Not only is it important to find upper limits on as many of these terms as possible, but the limits should also be tight as well as ``clean", i.e., the hypothetical signal for Lorentz violations should be understood quantitatively, e.g. in terms of Lorentz tensors the SME. Excluding systematic cancellations between the signals that are searched for and obtaining individual limits for each of them is possible by a detailed understanding of the physics underlying the experimental tests of Lorentz invariance.

Experimental tests that use optical or microwave cavities are based on a measurement of the resonance frequencies 
\be\label{fundamental}
\omega=2\pi\frac{mc}{2nL}
\ee
($m$ is a constant mode number, $c$ is velocity of light parallel to the cavity axis, $n$ the index of refraction of the medium, and $L$ the cavity length) of a cavity, defined by the boundary conditions for standing waves. A Lorentz-violating shift in $c$ and/or $L$ and $n$ connected to a rotation or boost of the apparatus would lead to a measurable shift in $\omega$. The shift in $L$ is mainly due to Lorentz violation in the electron's equation of motion \cite{ResSME} (a contribution from the photonic sector \cite{TTSME} is usually negligible). The shift in $c$ is caused by Lorentz violation in electrodynamics \cite{KosteleckyMewes}. While this contains an influence of the relative dielectric permittivity $\epsilon_r=n^2$, the shift in $n$ itself has not been considered hitherto. Cavity experiments provide separately the only existing limits on several photon and electron terms. They are now  performed by several research groups worldwide \cite{BrilletHall,cavitytests,Lipa,MuellerMM,Wolfsme}.

This work improves the theoretical description of the influence of the electron sector. This theory  \cite{ResSME} uses general Bloch wave functions to describe the electrons within the solid. The properties of the electrons enter through six quantities given by material-specific details of the Bloch functions. Unfortunately, detailed wave-functions are difficult to obtain, so these six quantities had to be estimated by a simple model. Here, we eliminate them, so that the theory now needs no assumptions that go beyond the use of Bloch states. Thus, the theory is very generally applicable and accurate. 

Subsequently, we consider the shift in the index of refraction $n$. A motivation for such a study is that experiments with matter-filled cavities are currently among the most accurate cavity experiments. Also, it is necessary for a detailed understanding of proposed interferometer experiments to measure the remaining unbound photon parameter (i.e., $\tilde\kappa_{\rm tr}$ as defined in Ref. \cite{KosteleckyMewes}) \cite{CPT04} . We also find a Lorentz-violating birefringence in initially isotropic media. Using this theory and the accurate and detailed results from recent cavity tests with vacuum- \cite{Lipa,MuellerMM} and matter-filled \cite{Wolfsme} cavities, we will be able to find limits on previously unbound electron coefficients and improved limits on others. We find that certain cavity materials show a large sensitivity of $L$ and/or $n$ on Lorentz violation and propose a new experiment based on the Lorentz-violating birefringence in materials, that may significantly improve the limits on parameters usually measured in Michelson-Morley experiments.

The effects we derive in this work depend on the symmetric electron coeffcients $c_{\mu\nu}^e$ and nine of the photon coefficients $(k_F)_{\kappa\lambda\mu\nu}$. The relevance of coordinate and field definitions in the SME has been discussed in the literature. Loosely speaking, suitable definitions move parameters of one sector into the other sectors, so only differential effects have a physical significance. For example, nine symmetric proton coefficients of $c^p_{\mu\nu}$ can be set to zero (and moved to the photon and electron sector) by a particular choice of coordinate and field definitions \cite{ColladayKostelecky,KosteleckyMewes,ResSME,redefs}. For this work (following Ref. \cite{ResSME}), we adopt these definitions. This allows us to disregard these proton terms and makes the electron and photon coefficients relevant here independently meaningful. The neutron terms $c_{\mu\nu}^n$ are also eliminated by that definition under the assumption that they are equal to the corresponding proton terms.

For the effects considered here, the nuclei contribute only by their mass and charge. This restricts the number of possible other coefficients from the proton and neutron sector, that cannot be eliminated by coordinate choices. For example, for non spin-polarized (i.e., not ferromagnetic) solids, any influence of terms that enter the non-relativistic Hamiltonian \cite{clockcomparison} proportional to a spin is removed (including the $\tilde b_J$ and some combinations of the $d-, g-,$ and $H-$ type coefficients for neutrons, protons, and electrons). A possible shift in the mass of neutrons and protons has been excluded to $10^{-27\ldots -32}$\,GeV \cite{clockcomparison,Bear}, so that we can ignore it for this work. We work to first order in the Lorentz-violating parameters throughout.

This manuscript is organized as follows: Sec. \ref{calc} describes the calculation for vacuum cavities, Sec. \ref{n} considers the index of refraction. In Sec. \ref{exp}, limits on Lorentz violation are derived from experiments and future experiments are discussed. Appendix \ref{largedispersion} discusses some cases of large dispersion ; Appendix \ref{LVB} constructs an explicit quantum-mechanical model of lattice vibrations with Lorentz violation.

\section{Vacuum cavities}\label{calc}

\subsection{Ansatz}

We start by sketching the theory given in \cite{ResSME}: The starting point is the nonrelativistic Hamiltonian $h=h_0+\delta h$ of a free electron, where $h_0$ is the usual free-particle Hamiltonian, and $\delta h$ a Lorentz-violating correction. The term of interest here corresponds to a modification of the kinetic energy of the electron \cite{clockcomparison}:
\begin{equation}\label{deltap}
\delta h = (\delta_{jk}+2E'_{jk})\frac{p_jp_k}{2m}+\ldots\,,
\end{equation}
where $m$ is the mass of the electron, $p_i$ the momentum components, and the dots denote terms that are left out here, as they have no consequences for (non spin-polarized) solids \cite{ResSME}. The $3\times 3$ matrix
\begin{equation}
E'_{jk}=-c_{jk}-\frac12c_{00}\delta_{jk}
\end{equation}
is given by the SME Lorentz tensor $c_{\mu\nu}$ for the electron (the theory can be generalized for spin-polarized solids) \cite{ResSME}. The idea is that the electrons inside crystals have a nonzero expectation value $\left<p_jp_k\right>$, which is a function of the geometry of the lattice. Therefore, Lorentz violation will cause a geometry change (``strain") of the crystal, which, in turn, leads to a measurable shift in the resonance frequency $\omega$ of a cavity made of the crystal. 

Strain is conventionally expressed by the strain tensor $e_{ij}$ \cite{Woan}. For $i=j$, it represents the relative change of length in $x_i$-direction, and for $i \neq j$, it represents the change of the right angle between lines originally pointing in $x_i$ and $x_j$ direction. Because a violation of Lorentz invariance is certainly tiny, it is sufficient to consider the linear terms in the Lorentz-violating coefficients $E'_{jk}$. A general linear relationship between the Lorentz violating $E'_{jk}$ and strain is given by 
\begin{equation}\label{edctensor}
e_{dc} = \mathcal B_{dcpj} E'_{pj}
\end{equation}
with a 'sensitivity tensor' $\mathcal B_{dcpj}$. It can be taken as symmetric in the first and last index pair ; symmetry under exchange of these pairs will hold only for some simple crystal geometries, like cubic. Thus, the tensor has at most 36 independent elements.

An explicit determination of the sensitivity tensor requires a theoretical model of electrons in solids. According to the Bloch theorem, electronic states can be described by Bloch wave functions (\cite{Ashcroft}, pp. 133-141). In \cite{ResSME}, these are used to determine the expectation value $\left<p_ip_j\right>$ contained in Eq. (\ref{deltap}) ; the corresponding strain is calculated using elasticity theory. As a result, the sensitivity tensor
\begin{equation}\label{bdefinition}
\mathcal B_{dcjp} =  \mu_{dcmp} \kappa_{mj}+\mu_{dcmj} \kappa_{mp}\,,
\end{equation}
where
\begin{equation}\label{kappadef}
\kappa_{mj}= \frac{N_{e,u}  \hbar^2}{m |\mbox{det}(l_{ij})|} k_{ml} k_{jk}\oxi_{lk} \,,
\end{equation}
can be calculated. $N_{e,u}$ is the number of valence electrons per unit cell, $|$det$(l_{ij})|$ is the volume of the unit cell expressed by the determinant of the matrix of the primitive direct lattice vectors, $k_{ml}$ is the matrix containing the primitive reciprocal lattice vectors, and $\mu_{dcmp}$ is  the elastic compliance tensor, the inverse of the elasticity tensor $\lambda_{dcmp}$ \cite{Woan}. The symmetric $3\times 3$ matrix $\oxi_{lk}$ is given by the Fourier coefficients of the Bloch wave functions. Its six parameters are unknown at this stage and had to be determined from a simple model that leads to $\oxi_{lk} \sim \gamma \delta_{lk}$, where $\gamma$ is a material dependent constant determined by the kinetic energy of the electrons. 

\subsection{Eliminating the unknowns}

The basic idea that eliminates the six $\oxi_{lk}$ is to use an alternative method to calculate the strain for the simple case of isotropic Lorentz violation $E'_{jk}\sim \delta_{jk}$ and to require that Eq. (\ref{edctensor}) yields the same result. If only the rotation invariant trace of the Lorentz-violation tensor is nonzero, 
\begin{equation}
E'_{ab}=\frac13 \tr(E)\delta_{ab}\, ,
\end{equation}
$\tr(E)$ can be absorbed into the conventional Hamiltonian by scaling the mass of the electron, which leads to a corresponding scaling of the Bohr radius $a_0=4\pi \epsilon_0
\hbar^2/(m_ee^2)$ \cite{H2SME}:
\begin{equation}
m_e \rightarrow m_e \left(1-\frac 23 \tr(E) \right), \quad 
a_0 \rightarrow a_0 \left(1+\frac 23 \tr(E) \right) \, .
\end{equation}
Thus, an isotropic expansion of the crystal will result, that is proportional to the scaling of the Bohr radius: $e_{dc}=2/3\delta_{dc} \tr(E)$. Equating this and Eq. (\ref{edctensor}), we obtain
\begin{equation}\label{isotrop}
2 \delta _{dc} = \mathcal B_{dcpj} \delta_{pj}
\end{equation}
or, inserting Eq. (\ref{bdefinition}),
\begin{equation}
\delta _{dc} = \mu_{dcmj} \kappa_{mj}\, .
\end{equation}
Multiplying by $\lambda_{dcab}$ using $\lambda_{dcab} \mu_{dcmj}=\delta_{am} \delta_{bj}$,
\begin{equation}\label{kappalambda}
\lambda_{ddab} = \kappa_{ab} \, .
\end{equation}
Since $\kappa_{ab}$ depends solely on material properties, this result is valid for general Lorentz violations and can now be reinserted into Eq. (\ref{bdefinition}):
\begin{equation}\label{sensnew}
\mathcal B_{dcjp} =
\mu_{dcmp}\lambda_{aamj}+\mu_{dcmj}\lambda_{aamp}\,.
\end{equation}
We now have expressed the sensitivity tensor by the elastic properties of the material, eliminating any assumptions on the Fourier coefficients contained in the Bloch wave function. At the same time, the reciprocal lattice vectors have dropped out. This is plausible, since ultimately the elastic properties are determined by the crystal structure (i.e., the lattice vectors) and the states of
the electrons. 

For later convenience, we express Eq. (\ref{edctensor}) as the equation 
\begin{equation}\label{esixvector}
\mathbf e=\mathcal B\cdot\mathbf E'
\end{equation}
between the six-vectors
\begin{equation}
\mathbf e=(e_{xx},e_{yy},e_{zz},e_{yz},e_{xz},e_{xy})
\end{equation}
and $\mathbf E'$, defined analogously. Therefore, we arrange the $B_{abcd}$ into a $6\times 6$ sensitivity matrix as defined in \cite{ResSME}. In what follows, we consider crystals of trigonal or higher symmetry (virtually all crystals presently used for cavities in precision measurements fall into this class, such as quartz, sapphire, and all isotropic materials): For cubical crystals, $\mathcal B$ has the same structure as given in \cite{TTSME}:
\begin{equation}\label{matrixcubic}
\mathcal B= \left( \begin{array}{cccccc} \mathcal B_{11} & \mathcal B_{12} & \mathcal B_{12} & 0 &
0 & 0
\\ \mathcal B_{12} & \mathcal B_{11} & \mathcal B_{12} &0 & 0& 0\\ \mathcal B_{12} & \mathcal
B_{12} & \mathcal B_{11} & 0& 0& 0\\ 0 & 0 & 0 & \mathcal B_{44} & 0& 0\\0 & 0 & 0 & 0 & \mathcal
B_{44} & 0
\\ 0 & 0 & 0 & 0 & 0 & \mathcal B_{44}
\end{array} \right) \, ,
\end{equation}
where
\begin{equation}\label{elementscubic}
\mathcal B_{11} = 2C_{11} S_1, \quad  \mathcal B_{12} = 2C_{12}S_1, \quad
\mathcal B_{44} = \frac 12 C_{44}(S_1+S_2).\nonumber
\end{equation}
The $S$s are the elements of the elasticity matrix $\mathbf S$, the $C$ those of the compliance matrix $\mathbf C=\mathbf S^{-1}$. We used 
\begin{equation}
S_1=S_{11}+S_{12}+S_{13} \, , \quad S_2 = 2S_{13}+S_{33} \, .
\end{equation}
[for cubical crystals, $S_{13}=S_{12}$.] For trigonal crystals, such as quartz and sapphire, we find
\begin{equation}\label{matrixtrigonal} 
\mathcal B = \left(
\begin{array}{cccccc}
  \mathcal B_{11} & \mathcal B_{12} & \mathcal B_{13} & \mathcal B_{14} & 0 & 0 \\
  \mathcal B_{12} & \mathcal B_{11} & \mathcal B_{13} & -\mathcal B_{14} & 0 & 0\\
  \mathcal B_{31} & \mathcal B_{31} & \mathcal B_{33} & 0 & 0 & 0 \\
  \mathcal B_{41} & -\mathcal B_{41} & 0 & \mathcal B_{44} & 0 & 0\\
  0 & 0 & 0 & 0 & \mathcal B_{44} & \frac 12 \mathcal B_{41} \\
  0 & 0 & 0 & 0 & \mathcal B_{14} & \mathcal B_{66}
\end{array} 
\right) \, .
\end{equation}
with $\mathcal B_{66} = ( \mathcal B_{11}-\mathcal B_{12})/4$. This has a simpler structure than Eq. (82) in \cite{ResSME}, as, for example, we now have
\begin{equation}
\mathcal B_{11}=\mathcal B_{22}, \quad \mathcal B_{21}=\mathcal B_{12}, \quad \mathcal B_{31}= \mathcal B_{32}, \quad \mathcal B_{41}=-\mathcal B_{42}
\end{equation} 
instead of the corresponding relations from \cite{ResSME}, that involved factors of 3. The new calculation, due to its elimination of the wave-function coefficients, corresponds to inserting a specific wave function for each crystal into the equations obtained in \cite{ResSME}. Since these have the full symmetry properties of the crystal, the result is simpler. In addition to Eqs. (\ref{elementscubic}), we find for trigonal crystals  
\begin{eqnarray}
\mathcal B_{13} &= &2C_{13}S_2\,, \quad \mathcal B_{14}  = 2C_{14}(S_1+S_2)\,, \nonumber \\
\mathcal B_{31} &= &2C_{13}S_1\,, \quad \mathcal B_{33} = 2C_{33}S_2, \quad \mathcal B_{41} = C_{14}S_1 \, . \quad
\end{eqnarray}

\subsection{Sensitivity for various materials}

Numerical values for some materials are given in Tab. \ref{numvalcub} and Tab. \ref{numvaltrig}, calculated from $\mathbf S$ as given in \cite{CRC}. For materials already considered in Ref. \cite{ResSME}, we find relatively minor changes, except for the case of sapphire, for which the old calculation gave comparatively low values. 

It is interesting to find materials with particularly high or low sensitivities (the latter for experiments predominatly sensitive to Lorentz violation in electrodynamics). Eq. (\ref{isotrop}) implies the six constraints 
\begin{equation}\label{sum13}
\sum_{i=1}^3 \mathcal B_{ij}=2 \, , \quad \sum_{i=4}^6 \mathcal B_{ij}=0
\end{equation} 
for $j=1,2,3$, that hold for any crystal geometry. However, the individual elements of $\mathcal B$ may vary for different materials. Materials presently in use include fused quartz \cite{BrilletHall}, niobium \cite{cavitytests}, and sapphire \cite{MuellerMM,Wolfsme}. Among these, niobium has the highest sensitivity coefficients. Calcium fluoride and other ionic materials have been considered for cavities in Ref. \cite{Bra01}.

\begin{table}
\caption{Sensitivity coefficients for isotropic materials, to be inserted into Eq. (\ref{matrixcubic}).  \label{numvalcub}}
\begin{tabular}{lrrrlrrr}\hline
Mat. & $\mathcal B_{11}$ & $\mathcal B_{12}$ & $\mathcal B_{44}$ & Mat. & $\mathcal B_{11}$ & $\mathcal B_{12}$ & $\mathcal B_{44}$ \\ \hline 
Al & 7.21 & -2.61 &	8.04 & Al$_2$O$_3$\footnote{fused sapphire} & 3.47 & -0.74 &  5.61 \\ 
Au & 24.13	& -11.06	& 12.34 & BaF$_2$ & 5.30 &	-1.65	& 6.82 \\ 
C\footnote{diamond; elastic coefficients form \cite{Kittel}} & 3.53 & -0.26 & 2.30 & CaF$_2$ & 3.46	& -0.73	& 3.00 \\
Fe & 8.51	& -3.26	& 4.36 & GaAs & 5.30 &-1.65	& 3.81 \\ 
Ge & 4.41	& -1.20	& 3.37 & Ir & 4.86	& -1.43 &	4.16 \\ 
Mo & 4.06	& -1.03 &	7.14 & Nb & 6.80	& -2.40 &	17.9 \\ 
Pb & 25.03 &-11.51	& 8.96 & Pt & 12.45 &	-5.22 &	11.09 \\ 
Si & 4.51 &	-1.26	& 3.69  & SiO$_2$\footnote{fused quartz} & 2.64 & -0.32 & 3.95 \\ 
W & 4.57 &	-1.29 &	5.79 & ZrC & 3.06 &	-0.53	& 4.20 \\ \hline
\end{tabular}
\end{table}

\begin{table}
\caption{Sensitivity coefficients for trigonal materials, to be inserted into Eq. (\ref{matrixtrigonal}).  \label{numvaltrig}}
\begin{tabular}{lrrrrrrrr}\hline
Mat. & $\mathcal B_{11}$ & $\mathcal B_{12}$ & $\mathcal B_{13}$ &
$\mathcal B_{14}$ & $\mathcal B_{31}$ & $ \mathcal B_{33}$ &
$\mathcal B_{41}$ & $\mathcal B_{44}$ \\ \hline 
Al$_2$O$_3$ & 3.58 & -1.05 & -0.53 & 0.014 &	-0.57 & 3.14 & 0.004 & 	5.08  \\ AlPO$_4$ & 5.95 & 	0.57 & -4.52 & 0.069 & -3.38 & 8.76 & 0.015 & 10.34
\\ CaCO$_3$ & 5.82 &	-1.97 &	-1.86 &	8.60	& -2.46	& 6.93 &	2.45	& 9.64\\
Fe$_2$O$_3$ & 2.76 &	-0.64	& -0.12	& 0.90	& -0.14	& 2.29	& 0.25	& 3.40 \\SiO$_2$ & 2.70	& -0.38	& -0.32	& 2.13	 & -0.26	& 2.52	& 0.48	& 2.35 \\  
\hline
\end{tabular}
\end{table}

Amongst the trigonal crystals in Tab. \ref{numvaltrig}, calcite and aluminum phosphate have the largest coefficients, about two times as large as for other trigonal materials. Calcite also has the larges $\mathcal B_{41}$, which may be important for a future measurement of $c_{0J}$ components of the Lorentz violation tensor, as described in the outlook. For cubic crystals (Tab. \ref{numvalcub}), the sensitivity of a typical optical cavity is given by $\mathcal B=\mathcal B_{11}-\mathcal B_{12}$, see Tab. \ref{signal}. For the materials presently in use, $3< \mathcal B < 10$. Silicon, with $\mathcal B=5.77$, is available in ultra-pure specimens and has a zero crossing of the thermal expansion coefficient. It has therefore been considered for future tests \cite{space}. The same might hold for other semiconductors, such as Ge and GaAs.  For gold, $\mathcal B=35.19$ is an order of magnitude higher than for most materials ; its high electric conductivity is beneficial for making microwave cavities. Lead, with similarly high coefficients, is a superconductor. A high sensitivity is found for materials whose elastic coefficients $S_{11}$ and $S_{12}$ are close to each other. 

\section{Matter-filled cavities}\label{n}

\subsection{The index of refraction in Lorentz tests}

For matter-filled cavity oscillators, the Lorentz violating shift in the index of refraction $n$ has to be considered in addition to the shifts in $c$ and $L$. We do not consider birefringence (however, see Sec. \ref {photonparam}), but for birefringent media our calculation will hold for any of the ordinary or extraordinary rays. To some extent, we shall follow Ref. \cite{Bra01} who calculated the change in the index of refraction due to a hypothetical time-dependence of the fundamental constants $\alpha$ and $m_e/m_p$.

The index of refraction $n(\omega,\pi_a)$ is a function of the frequency $\omega$ of the radiation and of the Lorentz-violation parameters that, for now, we denote $\pi_a$, $a=1,2,\ldots$. Examples are the electron parameters $c_{\mu\nu}$ and the photon parameters  $(\kappa_{DE})_{ij}$. The index of the refraction may therefore change both due to changes in $\delta \omega$ of the frequency as well as due to the $\pi_a$:
\begin{equation}
\delta n=\frac{\partial n}{\partial \omega}\delta \omega+\sum_a \frac{\partial n}{\partial \pi_a}\pi_a\,.
\end{equation}
Inserting into Eq. (\ref{fundamental}) and solving for the relative frequency change, 
\begin{equation}\label{totaleffect}
\frac{\delta \omega}{\omega}=\frac{1}{1+\bar n}\left[\frac{\delta c}{c}-\frac{\delta L}{L}-\sum_a\left(\frac1n\frac{\partial n}{\partial \pi_a}\right)\pi_a\right]\,
\end{equation}
where we denoted the normalized dispersion
\begin{equation}
\bar n=\frac \omega n\frac{\partial n}{\partial\omega}\,.
\end{equation} 
In the local field model \cite{Panofsky}, the index of refraction can be related to the susceptibility $\chi(\omega)$ by the Clausius-Mosotti equation \cite{CMremark}
\begin{equation}\label{clausiusmosotti}
\frac{n^2(\omega,\pi_a)-1}{n^2(\omega,\pi_a)+2}=\frac{\chi(\omega,\pi_a)}{3}\,.
\end{equation} 
Differentiating with respect to $\omega$ or $\pi_a$ on both sides, we find
\begin{equation}\label{nderiv}
\frac 1n\frac{\partial n}{\partial (\omega,\pi_a)}=\frac{(n^2+2)^2}{18n^2}\frac{\partial \chi}{\partial (\omega,\pi_a)}\,.
\end{equation}
We shall assume an ionic crystal, where the dielectric susceptibility is due to both electronic transitions, with typical resonance frequencies $\Omega$ in the ultraviolet spectral range, and optical phonon modes at infrared frequencies. Examples for such crystals are sapphire, calcium fluoride, calcium chloride, etc. In the transparency ranges, the susceptibility has the following approximate dispersion relation \cite{Palik}:
\begin{equation}\label{dispersion}
\chi = \frac{e^2}{\epsilon_0}\sum_{i}\frac{N_i}{M_i}\frac{f_i}{\Omega_i^2-\omega^2}
\end{equation}
where $\epsilon_0$ is the vacuum dielectric constant. $N_i$ is the density, $M_i$ the effective reduced mass, $f_i$ the oscillator strength, and $\Omega_i$ the resonance frequency of the $i$th optical mode. By calculating the derivative with respect to $\pi_a$, we obtain
\begin{eqnarray}\label{chideriv}
\frac{\partial\chi}{\partial\pi_a}&=&\frac{e^2}{\epsilon_0}\sum N_i\frac{f_i}{\Omega_i^2-\omega^2}\left(\frac{1}{N_i}\frac{\partial N_i}{\partial\pi_a}-\frac{1}{M_i}\frac{\partial M_i}{\partial\pi_a} \right. \nonumber \\ & & \left. +\frac{1}{f_i}\frac{\partial f_i}{\partial\pi_a}\right)+\sum_i\frac{\partial \chi}{\partial\Omega_i}\frac{\partial \Omega_i}{\partial\pi_a}\,.
\end{eqnarray} 
The oscillator strengths are dimensionless numbers of order unity for allowed transitions. According to exact sum rules \cite{Palik}, they sum up to unity, so an influence of Lorentz violation common to all the $f_i$ is excluded. A possible shift of the individual $f_i$ is still possible, but will be suppressed relative to the influences of the other quantities entering $\chi$. We neglect it in what follows.

A change in the masses $M_i$ would be connected to Lorentz violation in the hadron sector. As discussed in the introduction, such a change has been ruled out to extremely high precision \cite{clockcomparison,Bear,atoms}, so we may assume the masses are constant.

The densities $N_i$ are given by the inverse distances of the particles in the lattice, i.e., by the primitive translations of the reciprocal lattice. Lorentz violation leads to a geometry change of the lattice and thus to a change of the densities that is common for all $i$, i.e., 
\begin{equation}
\frac{1}{N_i}\frac{\partial N_i}{\partial \pi_a}=\frac{1}{N}\frac{\partial N}{\partial \pi_a} \,.
\end{equation}
By substituting $\chi$ by Eq. (\ref{clausiusmosotti}) and using Eq. (\ref{nderiv}), we obtain the relation
\begin{equation}\label{nderiv1}
\frac 1n\sum_a\frac{\partial n}{\partial\pi_a}\pi_a=\eta+d
\end{equation}
with
\begin{eqnarray}
\eta &=&\frac \beta 2 \sum_a \frac 1N \frac{\partial N}{\partial\pi_a}\pi_a\,,\nonumber \\ d &=&\frac{(n^2+2)^2}{18n^2}\sum_i\frac{\partial \chi}{\partial\Omega_i}\sum_a\frac{\partial\Omega_i}{\partial\pi_a}\pi_a\,,\nonumber \\ \beta &\equiv& \frac{(n^2+2)(n^2-1)}{3n^2}\,.
\end{eqnarray}
Let us call $\eta$ the density term and $d$ the dispersion term. In what follows, we insert the electron coefficients $c_{\mu\nu}$ for the $\pi_a$. The influence of the photon parameters is discussed separately.

\subsection{The density term $\eta$}

The density term can be determined using the theory for the geometry change of solids developed in \cite{ResSME} and above. The density of the material can be written as $N=N_0(1-\tr(e_{jk}))$, where $N_0$ is a constant that does not depend on Lorentz violation, and the tensor $e_{jk}$ expresses the strain due to Lorentz violation, given by Eq. (\ref{esixvector}) as a function of the Lorentz-violating parameters $E'_{jk}$ and the sensitivity coefficients of $\mathcal B$. Exact relations between the coefficients of $\mathcal B$ derived above allow to further simplify the density term. For a crystal of trigonal or higher symmetry, 
\begin{equation}\label{tre}
\tr(e)=(\mathcal B_{11}+\mathcal B_{12}+\mathcal B_{31})(E'_{xx}+E'_{yy})+(2\mathcal B_{13}+\mathcal B_{33})E'_{zz}\,.
\end{equation} 
Using the abbreviation 
\begin{equation}
E'_3=E'_{xx}+E'_{yy}-2E'_{zz}
\end{equation}
and Eq. (\ref{sum13}), we obtain 
\begin{equation}
\tr(e)=2\tr(E')+(\mathcal B_{31}-\mathcal B_{13}) E'_3\,.
\end{equation} 
We can thus rewrite the density term as
\begin{equation}\label{densityterm}
\eta=\frac \beta 2[2\tr(E')+(\mathcal B_{31}-\mathcal B_{13}) E'_3]\,.
\end{equation}

\subsection{The dispersion term}

\subsubsection{Optical frequencies}

Optical frequencies $\omega$ lie between the infrared (ir) resonance frequencies ($\sim 2\pi\times 3\times 10^{13}$\,Hz, for example) of the optical phonon modes and the electronic resonances in the ultraviolet (uv) spectral range (at, say, $2\pi\times 3\times 10^{15}$\,Hz). To estimate the relative influence of these resonances, we consider one resonance $\Omega_{\rm ir}$ in the infrared and one at $\Omega_{\rm uv}$ in the uv:
\begin{equation}\label{chiopt}
\chi = \frac{e^2}{\epsilon_0} \left(\frac{N_{\rm ir}}{M_{\rm ir}}\frac{f_{\rm ir}}{\Omega_{\rm ir}^2-\omega^2}+\frac{N_{\rm uv}}{M_{\rm uv}}\frac{f_{\rm uv}}{\Omega_{\rm uv}^2-\omega^2}\right)\,,
\end{equation}
We calculate  
\begin{eqnarray}\label{dispcomplete}
\frac{\partial \chi}{\partial \Omega_i}\frac{\partial \Omega_i}{\partial \pi_a} &=&-2\frac{e^2}{\epsilon_0} \left[\frac{N_{\rm ir}}{M_{\rm ir}}\frac{f_{\rm ir}\Omega^2_{\rm ir}}{(\Omega_{\rm ir}^2-\omega^2)^2}\left(\frac{1}{\Omega_{\rm ir}}\frac{\partial \Omega_{\rm ir}}{\partial \pi_a}\right)\right. \nonumber \\ & & \left.+\frac{N_{\rm uv}}{M_{\rm uv}}\frac{f_{\rm uv}\Omega_{\rm uv}^2}{(\Omega_{\rm uv}^2-\omega^2)^2}\left(\frac{1}{\Omega_{\rm uv}}\frac{\partial\Omega_{\rm uv}}{\partial \pi_a}\right)\right]\,.\quad
\end{eqnarray}
This shows that the influence of the infrared resonance is suppressed by a factor of about $(\Omega_{\rm ir}/\Omega_{\rm uv})^2\sim 10^4$. We will therefore neglect it and ascribe the whole susceptibility $\chi(\omega)$ at optical frequencies to the uv resonance. This is a good approximation, since the contribution of the ir mode to the susceptibility at optical frequencies is just a few \% (it is actually negative, compare to Eq. (\ref{dispersion})). Using Eqs. (\ref{clausiusmosotti},\ref{nderiv}), we can rewrite the dispersion term as
\be\label{dispsingle}
d=-\beta \frac{\Omega_{\rm uv}^2}{\Omega_{\rm uv}^2-\omega^2}\sum_a\left(\frac{1}{\Omega_{\rm uv}}\frac{\partial \Omega_{\rm uv}}{\partial\pi_a}\right)\pi_a \,,
\ee
or, additionally using $\omega \ll \Omega_{\rm uv}$, 
\begin{equation}\label{dispersionoptical}
d=-\beta\sum_a\left(\frac{1}{\Omega_{\rm uv}}\frac{\partial \Omega_{\rm uv}}{\partial\pi_a}\right)\pi_a \,.
\end{equation}
This equation allows to calculate the Lorentz-violating shift in $n$ from the Lorentz-violating shift in the atomic or molecular resonance frequency $\Omega_{\rm uv}$. 

As far as the atomic resonances are well described by a tight-binding model \cite{Harrison1,Harrison2}, some of the existing theoretical and experimental studies of Lorentz violating effects in atoms \cite{atoms,clockcomparison,H2SME} can be applied: The dependence on the angular momentum quantum numbers of the electronic state can be derived by application of the Wigner-Eckart theorem, i.e.,  the shift $\delta \Omega_{\rm uv}/\Omega_{\rm uv}$ is a function of the angular momentum quantum numbers of the valence electrons. 

\subsubsection{Infrared and microwave frequencies}

At ir or lower frequencies, the influence of the ir resonance in Eq. (\ref{dispcomplete}) outweighs the one of the atomic uv resonances, because the oscillator strengths of the ir resonances (optical phonon modes) are higher \cite{Palik}. We may then ascribe the whole susceptibility to the ir resonance, and obtain for the dispersion term
\be\label{dispir}
d=-\beta\frac{\Omega_{\rm ir}^2}{\Omega_{\rm ir}^2-\omega^2}\sum_a\left(\frac{1}{\Omega_{\rm ir}}\frac{\partial \Omega_{\rm ir}}{\partial\pi_a}\right)\pi_a \,.
\ee
The case of multiple ir resonances is discussed below.  

For an illustration, consider a lattice consisting of dissimilar atoms with masses $M_1$ and $M_2$ within each unit cell separated by a distance $a$ and connected with springs having a force constant $\kappa$. The frequency $\Omega(q)$ of a phonon with a wave number $q$ splits into acoustic and optical modes. The vibration spectrum is given in textbooks by \cite{Harrison1}
\begin{eqnarray}
\Omega^2(q)&=&\frac{2\kappa}{M_1M_2}\left(\frac{M_1+M_2}{2}\right. \nonumber \\ && \left. \pm\sqrt{\frac{(M_1-M_2)^2}{4}+M_1M_2\cos^2 qa}\right)\,.
\end{eqnarray}
The $+$ and $-$ signs correspond to the optical and acoustical modes, respectively. In the optical modes, the two atoms oscillate in paraphase. If they have different charges, this leads to an oscillating dipole moment that interacts with electromagnetic waves. Since the speed of sound in solids is much lower than $c$, we may consider the case $qa\ll 1$ for these phonons, obtaining 
\be
\Omega\approx \Omega(0)=\sqrt{\frac{\kappa}{\bar M}}\,,
\ee
for the resonance frequency, where $\bar M=M_1M_2/(M_1+M_2)$ denotes the reduced mass. (In a three-dimensional lattice in the most general case, there will be two transversal modes and one longitudinal mode that correspond to different effective spring constants and masses.) The masses are mainly due to the nuclei and will thus not appreciably change due to Lorentz violation in the photon and electron sector. An influence of Lorentz violation comes in through the spring constant.

The frequencies of the long-wavelength longitudinal modes in ionic crystals are given by the ion-plasma frequency \cite{Harrison1} 
\be
\Omega_{\rm IP}=\sqrt{\frac{4\pi Z^2e^2}{\bar M V}}
\ee
[$V$ is the atomic volume, $Z$ the valence of the ions]. Under the influence of Lorentz violation, the volume $V=V_0(1+\tr(e_{jk}))$ can be expressed by the strain tensor considered above in this paper and a constant $V_0$ that does not depend on Lorentz violation. The other quantities in $\Omega_{\rm IP}$ do not depend on Lorentz violation. Therefore, 
\be
\left(\frac{1}{\Omega_{\rm IP}}\frac{\partial \Omega_{\rm IP}}{\partial\pi_a}\right)\pi_a= -\frac 12 \tr(e_{jk})\,.
\ee
Using $\Omega_{\rm IP}$ for $\Omega_{\rm ir}$ in the dispersion term for infrared or lower frequencies, Eq. (\ref{dispir}), we obtain
\begin{eqnarray}\label{irres}
d&=&\frac \beta 2 \frac{\Omega_{\rm ir}^2}{\Omega_{\rm ir}^2-\omega^2}\tr(e_{jk}) \nonumber \\&=& \frac \beta 2 \frac{\Omega_{\rm ir}^2}{\Omega_{\rm ir}^2-\omega^2}[2\tr(E')+(\mathcal B_{31}-\mathcal B_{13}) E'_3]\,. 
\end{eqnarray}
The second line, which holds for a crystal of trigonal or higher symmetry, has been derived by use of Eq. (\ref{tre}). Generalizing for multiple ir resonances whose frequencies $\Omega_{{\rm ir,} i}$, ($i=1,2,\ldots$) are all proportional to $\Omega_{\rm IP}$, we obtain  
\be\label{multirres}
d= \frac \beta 2 [2\tr(E')+(\mathcal B_{31}-\mathcal B_{13}) E'_3] \sum_{i}K_i\frac{\Omega^2_{{\rm ir,} i}}{\Omega^2_{{\rm ir,} i}-\omega^2}  
\ee
where the factors $K_i$ ($\sum_i K_i=1$) describe the relative strengths as given by $f_i, M_i$, and $N_i$ according to Eq. (\ref{dispersion}). For $\omega\ll\Omega$ (e.g., for microwave frequencies), we obtain
\be 
d=\eta
\ee
for a single or for multiple ir resonances.

\subsection{Total effect}

According to Eq. \ref{totaleffect}, the sensitivity of the cavity resonance frequency $\omega$ to Lorentz violation is proportional to the factor $1/(1+\bar n)$. With the exception of frequencies $\omega$ near one of the resonances of the medium (which we discuss in Appendix A) for usual materials in the transparency range the normalized dispersion $|\bar n| \lesssim 0.01$. In the following, we can thus assume that $1/(1+\bar n)\approx 1$. Thus, the length change $\delta L/L$ and the change $\delta c/c$ affect the cavity resonance frequency unattenuated, as in a vacuum cavity.

The density term $\eta$ has been calculated from our results for the geometry change of crystals. The dispersion term $d$ has been evaluated for the case of infrared and microwave frequencies. For the simplest case, microwave frequencies, the dispersion term turns out to be equal to the density term. In this case, the total change of the index of refraction due to Lorentz violation can be written as
\be\label{totallowfreq}
\frac 1n \sum_a \frac{\partial n}{\partial \pi_a}\pi_a=d+\eta=2 \beta \tr{E'}+\bar\beta E'_3
\ee
where 
\be 
\bar\beta=\beta(\mathcal B_{31}-\mathcal B_{13})\,.
\ee
Since the trace of the Lorentz-violating quantity is a constant under spatial rotations, it will not show up in the signal of a typical Lorentz test, that searches for a signal that varies with spatial rotation (see below). The influence of $E'_3$ is given by $(\mathcal B_{31}-\mathcal B_{13})$ which is zero for isotropic materials ; for such materials, therefore, the density term does not contribute to the signal of experiments. Tab. \ref{barbeta} shows $\bar\beta$ for some materials, calculated from Tab. \ref{numvaltrig} and values $\epsilon$ from \cite{CRC}. The indices $\bar \beta_{11}, \bar \beta_{22}$ etc. indicate the polarization of the radiation with respect to the crystal coordinates. The values of $\bar\beta$ differ greatly for different materials, from about $-0.1$ to about 2.5, mostly due to different values of $\mathcal B_{31}-\mathcal B_{13}$.

\begin{table}
\caption{Index-of-refraction change coefficient $\bar \beta$ for microwave frequencies. For values in italics, the polarization is not specified. \label{barbeta}}
\begin{tabular}{lrrrlrrr}\hline
Mat. & $\bar\beta_{11}$ & $\bar\beta_{22}$ & $\bar\beta_{33}$ & Mat. & $\bar\beta_{11}$ & $\bar\beta_{22}$ & $\bar\beta_{33}$ \\ \hline
Al$_2$O$_3$ & -0.12 & $=\bar\beta_{11}$ & -0.15 & CaCO$_3$ & -1.91 & -1.92 & -1.84 \\ Fe$_2$O$_3$ & & & {\em -0.042} & SiO$_2$ & 0.098 & 0.097 & 0.102 \\AlPO$_4$ & & & {\em 2.54} &&&& \\
\hline
\end{tabular}
\end{table}

The index-of refraction change at ir frequencies can be calculated from Eqs. (\ref{irres},\ref{multirres}). The data on the infrared resonance frequencies and their relative strengths $K_i$, as well as the index of refraction $n(\omega)$ itself, can be obtained from tables of Sellmeier coefficients (see, e.g., \cite{Palik}). 

The factor $\beta=(n^2-1)(n^2+2)/(3n^2)$ means that for large $n$ the relative change $\delta \omega/\omega$ in the cavity resonance frequency may be greater than the relative change $\delta \Omega_i/\Omega_{i}$. At microwave frequencies, for example, Al$_2$O$_3$ has $\beta_{33} \sim 4.1$ (however, $\bar\beta$ in this case is small due to the low value of the $\mathcal B_{13}-\mathcal B_{31}$); AlPO$_4$ has $\beta=2.24$ and $\bar\beta =2.54$. This is a consequence of the Clausius-Mosotti equation, which implies that for a large $n$, a small relative change in $\chi$ causes a large relative change in $n$. Certain ceramics materials intended for microwave dielectric resonator oscillators (DRO) have a dielectric constant $\epsilon_r=n^2$ of 60 and thus an even larger $\beta\sim 20$. 

The rise in the sensitivity will even become more pronounced at higher ir frequencies, where both $n^2$ as well as the factor $\Omega^2_{\rm ir}/(\Omega^2_{\rm ir}-\omega^2)$ in Eqs. (\ref{irres},\ref{multirres}) become larger. Very close to a resonance, however, a large normalized dispersion $\bar n$ eventually limits this rise of the sensitivity, as discussed in Appendix A.

\subsection{Influence of the photon parameters and Lorentz-violating birefringence}\label{photonparam}

The photon parameters will influence the ion-plasma frequency $\Omega_{\rm IP}$ directly through a modification of the Coulomb potential of the ions and indirectly through a change of the atomic volume $V$. 

The ion-plasma frequency $\Omega$ describes oscillations of the ions in the potential due to their Coulomb interaction. For small dilations $\delta \vec r$ from the equilibrium position, this potential is approximately harmonic. In the SME, the Coulomb potential $\Phi$ between two charges $e$ is given by \cite{KosteleckyMewes}
\be\label{coulombsme}
\Phi(\vec r)=\frac{ e^2}{4\pi r}\left(1+\frac 12 \frac{\vec r \kappa_{DE} \vec r}{r^2}\right)\,,
\ee
where $r\equiv |\vec r|$ and $\kappa_{DE}$ is a $3\times 3$ matrix specifying Lorentz violation in the photon sector from the SME. 

Let us study this influence to a crystal that is initially isotropic, using coordinates in which $\kappa_{DE}$ is diagonal with principal values $(\kappa_{DE})_{11},(\kappa_{DE})_{22}$, and $(\kappa_{DE})_{33}$. Elastic waves with a wavevector $\vec q$ can be described by the ansatz \cite{Harrison1} $\delta \vec r=\vec u_q e^{i \vec q\vec r}$. As the ions are charged by $Ze$ ($Z$ is the valence), the dilation of the ions due to the elastic wave changes the local charge-density, thus causing an electrostatic potential. The negative gradient of that potential gives the resulting force \cite{Harrison1} \be
F= -4\pi Z^2 e^2 \vec u_q/Ve^{i\vec q\vec r}\,,
\ee 
which acts back on the ions, causing oscillations with $\Omega_{\rm IP}$. In the SME, this force becomes dependent on the direction of the dilation, which is given by $\vec u_q$. For example, if $\vec u_q$ is parallel to the $x$-direction, then according to Eq. (\ref{coulombsme}) the force should be replaced by
\be
F=\frac{-4\pi Z^2 e^2 \vec u_q}{V}\left(1+\frac 14(\kappa_{DE})_{11}\right)e^{i\vec q\vec r}\,.
\ee
The ion-plasma resonance will thus split into three modes according to the direction of $\vec u_q$ with respect to the coordinates,
\be\label{omipchange}
(\Omega_{\rm IP})_{aa} = \Omega_{\rm IP}\left(1+\frac 14 (\kappa_{DE})_{aa}\right)\,,
\ee
where $aa$ (no summation convention) may be $11, 22$, or $33$. If the plasma oscillations are caused by an optical wave, the direction of $\vec u_q$ will be given by the polarization. Thus, an initially isotropic medium will become birefringent ; for a optical wave propagating into the $z$-direction, for example, there will be a difference 
\be 
\frac{\Delta n}{n} =\frac{\beta}{4} [(\kappa_{DE})_{11}-(\kappa_{DE})_{22}]
\ee
between the index of refraction for $x$ and $y$ polarized light. The consequences of this birefringence are as usual, e.g, the propagation speed will depend on the polarization, linearly polarized light not aligned with one of the principal axes will change its polarization while propagating etc.

In vacuum, the Lorentz violation in electrodynamics described by the matrix $\kappa_{DE}$ leads to a dependence of the speed of light on the direction of propagation, but not to birefringence (which is caused by other parameters, which are, however, ruled out to at least $2\times 10^{-32}$ by measuring the polarization of light from astronomical sources, see Ref. \cite{KosteleckyMewes}). In media, however, it causes birefringence through its influence on the resonances of the medium. 

In the model of ion-plasma oscillations, the electron parameters do not cause birefringence, as the model contains only the electrostatic interaction between the ions. A more realistic model of a cubic lattice (see Appendix B), however, predicts that the electron parameters cause birefringence as well: the matrix $\tilde\kappa_{DE}$ in Eq. (\ref{omipchange}) should be replaced by
\be
2\tilde\kappa_E\equiv 4E'_{zz}+3\frac{\alpha_3}{\alpha_1\eta}(\tilde\kappa_{DE})_{zz}\,,
\ee 
where $\alpha_1\approx -1.74 $ and $\alpha_3\approx -3.24$ are numerical factors determined by the crystal geometry.

The change of the atomic volume $V$ due to the photon parameters can be obtained from the material given in Ref. \cite{TTSME}. Therein, two parameters $a_\perp$ and $a_\|$ give the relative change of the length of a cavity. The model used therein assumes a cubic crystal. For the change of the volume, we obtain by summation over the relative changes of the length in $x,y,$ and $z$ direction: 
\be
V=V_0(a_\|+2a_\perp)\tr(\kappa_{DE})\,,
\ee
where $V_0$ does not depend on Lorentz violation and the $3\times 3$ matrix $\kappa_{DE}$ parametrizes Lorentz violation in the electrodynamic sector \cite{KosteleckyMewes}. The parameters have been calculated for some cubic crystals such as NaCl, LiF, and NaF. For these, $-0.3\leq a_\|+2a_\perp\leq -0.1$. Since for cubic crystals, the volume depends solely on the rotation invariant trace (as expected from symmetry arguments) no influence to the outcome of typical experiments will result. The parameters have also been estimated for anisotropic materials such as quartz ($a_\|+2a_\perp=-0.03$) and sapphire ($a_\|+2a_\perp=-0.01$) \cite{TTSME}. These values are small because the chemical bonds in these materials have a partial covalent character. Although the anisotropic parameters have not been considered, the smallness of the isotropic ones should carry over to them.  

The most general Maxwell equations that are linear and first order in the derivatives might contain further hypothetical Lorentz-violating photon terms that are not contained in the SME \cite{CNC}. As discussed in \cite{MLCPT04}, these terms do not cause an additional modification of the geometry of solids, and, based on the same arguments, no additional modification of $n$.

\section{Experiments}\label{exp}

\subsection{Limits from published data}

\begin{table*}
\caption{\label{signal} Signal components and experimental limits for the experiment by Brillet and Hall \cite{BrilletHall} (upper two lines) and M\"uller {\em et al.} \cite{MuellerMM} (lower two). $\mathcal B_q\simeq -2.96$ and $\mathcal B_s\simeq-3.71$ are combinations of material parameters $\mathcal B_{32}-\mathcal B_{33}$ for fused quartz and sapphire, respectively. $\chi_B \simeq 50^\circ$ is the colatitude of Boulder, Colorado, $\chi \simeq 42.3^\circ$ the one of Konstanz. We abbreviated $c_3=(c_{xx}+c_{yy}-2c_{zz})$, $c_2=c_{XX}-c_{YY}$, and analogous for $\tilde \kappa_{e-}$. As usual, $c_{(ab)}=\frac12 (c_{ab}+c_{ba})$.}
\begin{tabular}{lcrcr}\hline
Frequency & sine amplitude & limit$/10^{-15}$ & cos amplitude & limit$/10^{-15}$ \\ \hline
$2\omega_t$ & 0 & $\sim 200$ & $\frac{1}{12} \sin^2\chi_B \left[\frac 43 \mathcal B_qc_3  +(\tilde\kappa_{e-})^{zz}\right]$ & $\sim 200$ \\ $2\omega_t+2\oo$ & $\frac 14 \cos^4\frac{\chi_B} {2}\left[(\tilde\kappa_{e-})^{xy}-4\mathcal B_q c_{(xy)}\right] $ & 4 & $\frac18 \cos^4\frac{\chi_B}{2} \left[-4\mathcal B_qc_2 +(\tilde \kappa_{e-})_2\right]$ & 4 \\ \hline 

$\oo$ & $\frac12\sin2\chi[4\mathcal B_sc_{(YZ)}-(\tilde\kappa_{e-})^{YZ}]$ & $-1.8\pm4.5$ & $\frac12 \sin 2\chi[4\mathcal B_sc_{(XZ)}-(\tilde\kappa_{e-})^{XZ}]$ & $3.2\pm6.2$ \\ $2\oo$ & $-2\mathcal B_s\cos^2\chi c_{(XY)}+\frac12(1+\cos^2\chi)(\tilde\kappa_{e-})^{XY}$ & $1.3\pm2.3$ & $-\mathcal B_s\cos^2\chi c_2+\frac14(1+\cos^2\chi)(\tilde\kappa_{e-})_2$ & $3.4\pm1.9$ \\ \hline
\end{tabular}
\end{table*}

We derive limits from the experiments by Brillet and Hall \cite{BrilletHall}, who used a quartz cavity, and M\"uller {\em et al.} \cite{MuellerMM}, who used a crystalline sapphire cavity. 

As usual, we exploit the specific time-dependence imposed on the frequency shift of cavities on Earth by Earth's rotation and orbit ('signal'). The signal for Lorentz violation in electrodynamics has been calculated in \cite{KosteleckyMewes}, the one for Lorentz violation in the electron's equation of motion in  \cite{ResSME}. Throughout, we are using the conventions described there. $\Omega_\oplus$ is the sidereal angular frequency of Earth's orbit, $\omega_\oplus$ the one of Earth's rotation, and $\omega_t$ the angular frequency of the rotation of the turntable, as measured in the laboratory. The signal consists of sine and cosine components at various combinations of these frequencies. In the first two lines of Tab. \ref{signal}, we list the components measured by Brillet and Hall, those measured by M\"uller {\em et al.} are given in the last two lines. Additional signal components that include no contribution from the electronic sector are omitted here and are as given in \cite{MuellerMM}.  

Solving the signal components for the Lorentz-violation parameters (adding the errors in quadrature), we obtain the separate limits quoted in Tab. \ref{BHresults}. Some of the signal components cannot be separated into the electrodynamic and electronic contributions (as results for only two signal frequencies have been reported by Brillet and Hall). The error of these limits is dominated by the Brillet-Hall experiment. Three additional limits on the purely electrodynamic parameters $\tilde\kappa_{o+}$ are as stated in \cite{MuellerMM}. 

\begin{table}
\caption{Limits on photon and electron coefficients derived from Refs. \cite{BrilletHall,MuellerMM}. \label{BHresults}}
\begin{tabular}{lr}\hline
coefficient & limit$/10^{-15}$ \\ \hline
$(\tilde\kappa_{e-})^{XY}$ & $1.8\pm 12$ \\ 
$(\tilde\kappa_{e-})^{XX}-(\tilde\kappa_{e-})^{YY}$ & $15\pm39$ \\ 
$c_{(XY)}$ & $-0.16\pm3.0$ \\
$c_{XX}-c_{YY}$ & $-1.33 \pm 7.2$ \\
$c_{(YZ)}-0.067(\tilde\kappa_{e-})^{YZ}$ & $0.24\pm 0.61$ \\
$c_{(XZ)}+0.067(\tilde\kappa_{e-})^{XZ}$ & $-0.43\pm0.84$ \\
$|c_{XX}+c_{YY}-2c_{ZZ}-0.25(\tilde\kappa_{e-})^{ZZ}|$ & $\lesssim 1000$ \\ \hline
\end{tabular}
\end{table}

For a complete determination of the photon and electron coefficients, the experiment of M\"uller {\em et al.} has to be compared to an experiment that gives as many signal components, but uses dissimilar cavities. An experiment using a whispering-gallery type microwave sapphire cavity is reported by Wolf {\em et al.} \cite{Wolfsme}. In this cavity, the mode travels along the circumference of a cylinder, where the cylinder axis is parallel to the $c$ axis. 98\% of the electromagnetic field energy are confined within the material. The signal in this experiment contains contributions from the shift in $c$, treated in much detail therein, as well as shifts in $L$ and $n$. Since we have derived all these shifts, we are now in a position to analyze this experiment against the one of M\"uller {\em et al.}, to obtain separate limits on electron and photon coefficients.

The change of the circumference $L$ for this ring cavity is given by 
\begin{equation}
\frac{\delta L}{L} = \frac 12(e_{xx}+e_{yy})=\frac12(\mathcal B_{11}+\mathcal B_{12})(E'_{xx}+E'_{yy})+\mathcal B_{13}E'_{zz}\,,
\end{equation} 
or, using Eq. (\ref{sum13}), 
\begin{equation}
\frac{\delta L}{L}=\frac23 \tr(E')+\left(\frac13-\frac12\mathcal B_{13}\right)E'_3\,.
\end{equation}
The time dependence of this is calculated as described in \cite{ResSME}. For the change in the index of refraction, we are in the limiting case of low frequencies, so Eq. \ref{totallowfreq} applies. The total frequency change due to electron parameters is thus
\begin{eqnarray}
\left.\frac{\delta\omega}{\omega}\right|_e =-\left(\frac13-\frac12\mathcal B_{13}\right)E'_3 +\bar \beta E'_3 \,.
\end{eqnarray}
We disregarded the terms proportional $\tr(E')$, as they will not cause time-dependent signals due to their rotation invariance. The time-dependence of the signal in the laboratory frame is calculated from the material given in \cite{ResSME}, see Tab. \ref{signal}. The strong dissimilarity of the cavities in the two experiments (the ring mode normal to the $c$ axis of Wolf {\em et al.} versus a linear mode parallel to the $c$ axis used by M\"uller {\em et al.}) leads to a strong dissimilarity of the material coefficients entering the signal $\mathcal B_w=+0.60$ versus $\mathcal B_s=-3.71$, although both experiments use sapphire cavities. This contributes to a high accuracy for the Lorentz violation parameters. For $\bar \beta$, we insert $\bar\beta_{33}$ for sapphire from Tab. \ref{barbeta}, according to the polarization actually used.  

The limits on the photon parameters $\tilde\kappa_{o+}$ are as quoted in \cite{MuellerMM,Wolfsme}. We now have a more complete set of limits on the total 19 photon and 10 electron coefficients. All coefficients measurable in present experiments could be disentangled. 

\begin{table}
\caption{Signal components for Lorentz violation in the electron sector for the experiment of Ref. \cite{Wolfsme}, and results reported therein. $\mathcal B_W=\frac 13-\frac 12\mathcal B_{13}\simeq 0.598$ is a combination of sapphire coefficients. Paris happens to be located at the same colatitude $\chi\simeq 42.3^\circ$ as Konstanz. The error here is the geometric sum of the statistic and systematic error given in \cite{Wolfsme}. \label{wolfresults}}
\begin{tabular}{lrr}\hline
component & signal & limit$/10^{-15}$ \\ \hline
$\sin \oo\To$ & $3(\mathcal B_W-\beta)c_{(YZ)}\sin 2\chi$ & $0.24\pm 0.59$ \\
$\cos\oo\To$ & $3(\mathcal B_W-\beta)c_{(XZ)}\sin 2\chi$ & $1.4\pm 0.59$ \\
$\sin 2\oo\To$ & $3(\mathcal B_W-\beta)c_{(XY)}\sin^2\chi$ & $1.1\pm 0.44$ \\
$\cos 2\oo\To$ & $\frac 32(\mathcal B_W-\beta)(c_{XX}-c_{YY})\sin^2\chi$ & $0.31\pm 0.44$ \\ \hline
\end{tabular}
\end{table}

\begin{table}
\caption{Limits on photon and electron parameters derived from Refs. \cite{MuellerMM,Wolfsme}. \label{limitswolfmueller}}
\begin{tabular}{lr}\hline
coefficient & limit $/10^{-15}$ \\ \hline
$(\tilde\kappa_{e-})^{YZ}$ & $0.52\pm 2.52$ \\
$(\tilde\kappa_{e-})^{XZ}$ & $-4.0\pm 3.3$ \\
$(\tilde\kappa_{e-})^{XY}$ & $-1.7\pm 1.6$ \\
$(\tilde\kappa_{e-})^{XX}-(\tilde\kappa_{e-})^{YY}$ & $2.8\pm 3.3$ \\ 
$c_{(YZ)}$ & $0.21\pm 0.46$ \\
$c_{(XZ)}$ & $-0.16\pm 0.63$ \\
$c_{(XY)}$ & $0.76\pm 0.35$ \\
$c_{XX}-c_{YY}$ & $1.15\pm 0.64$ \\ \hline
\end{tabular}
\end{table}

\subsection{New setups}\label{newsetups}

Here, we discuss experiments that exploit the results of this work in order to obtain enhanced sensitivity on Lorentz invariance violations.

By an appropriate choice of materials, large length change coefficients can be used to obtain a sensitivity that is higher than for the materials presently in use. For example, the coefficients are large for gold and lead, which may be useful for making microwave cavities, as gold exhibits a high electrical conductivity and lead superconductivity at low temperatures. The sensitivity of the index of refraction is enhanced relative to the one of the material resonance frequencies by factors of about 1-20 for materials with a large $n$ between 1 and 60. (See Appendix A for the potential of obtaining even larger sensitivity by operating near a resonance frequency of the medium.)

For an illustration, consider a measurement of the symmetric parity-odd electron coefficients $c_{(0J)}$, that have not yet been measured in laboratory experiments \cite{BaileyKostelecky}. While these coefficients do not directly affect the resonance frequencies of cavities, the Lorentz transformations due to Earth's orbital motion with the velocity $\beta_\oplus\sim10^{-4}$ makes them contribute to the measurable coefficients in the laboratory frame. Thereby, they cause a contribution of the order of $\beta_\oplus$ to the signal for experiments using anisotropic cavity materials (for isotropic materials, they can only contribute to order $\beta_\oplus^2$ or higher) \cite{ResSME}. For example, using Fabry-Perot cavities out of a trigonal material, with the mode orthogonal to the $c$ axis, they can be measured to order $\beta_\oplus\mathcal B_{14}$. The detailed signal components are given in \cite{ResSME}. Calcite has $\mathcal B_{14}=8.60$, sapphire 0.014 (Tab. \ref{numvaltrig}) ; a turntable experiment using a cavity made out of calcite, in which the Lorentz-violating frequency changes would be bounded to a level of $10^{-15}$ would provide bounds on the $c_{(0J)}$ of the order of $10^{-12}$. For those coefficients, this would be 600 times better than a comparable experiment using a sapphire cavity. 

\subsection{Michelson-Morley type experiment based on birefringence in matter}\label{birefringenceMM}

A new class of experiments would exploit the birefringence of isotropic materials that is caused by Lorentz violation, as predicted for infrared and microwave frequencies in Sec. \ref{photonparam} and Appendix \ref{LVB}. Since the birefringence depends on the same photon and electron coefficients $\tilde\kappa_{DE}, c_{\mu\nu}$ in the laboratory frame as $c$ and $L$, a sensitive birefringence measurement could access the same coefficients as a conventional Michelson-Morley type experiment. 

The most sensitive way to measure birefringence of some material is to fabricate a cavity filled with the material and compare the resonance frequencies $\omega_{x,y}=2\pi m c/(2n_{x,y} L)$ of the $x$ and $y$ polarized modes (we neglect the length change and the change of $c$ for this description). One possible realization of this idea uses three lasers whose frequencies $\omega_{1,2,3}$ are stabilized to three adjacent longitudinal modes of the cavity, corresponding to mode numbers $m-1,m,$ and $m+1$, respectively. The lasers at $\omega_1$ and $\omega_3$ would have the same constant linear polarization, while the polarization of laser 2 is switched (or continuously changed) between the $x$ and $y$ directions, for example by use of a rotating half-wave plate, a Pockels cell, or a liquid crystal device \cite{Hobbs}. A comparison of the three laser frequencies can then be used to measure the birefringence $\Delta n=n_x-n_y$ while eliminating the common influence to $\omega_{1,2,3}$ given by the cavity length $L$, the average index of refraction $n=(n_x+n_y)/2$, and the speed of light $c$. 

In conventional cavity experiments, thermal and other fluctuations that affect the cavity length and the index of refraction limit the accuracy. In the experiment proposed here, the elimination of $L$ and $n$ suppresses these influences. 

The degree of suppression depends on the residual birefringence of the cavity material. Fine annealed glass, the best commercial grade, has less than 12\,nm residual birefringence in 100\,mm thickness \cite{Hobbs}, i.e., $\Delta n/n \lesssim 10^{-7}$. As usual, the rotation of the Earth or active rotation of the cavity can be used to separate a Lorentz violation from this initial offset. Another systematic effect is stress-induced birefringence ; most optical glasses have piezo-optic coefficients of $2\times 10^{-12}$\,Pa$^{-1}$ \cite{Hobbs}. For a typical cavity with dimensions (30\,mm)$^3$, weighing about 50g, the birefringence caused by its own weight at the center of the cavity would thus be about $5\times 10^{-10}$. This is slightly lower than the length change caused by the weight, since the elastic modulus of such glasses is of the order $2\times 10^{11}$\,Pa \cite{CRC}. In usual cavity experiments, any change of the optical path length $nL$ inside the cavity caused by gravitational and other distortion would adversely influence the signal. In the proposed experiment, however, the optical path length drops out. Furthermore, the cavity itself can remain stationary while the polarization of one interrogating laser is rotated: if the orientation of the polarization is not correlated to the one of the cavity (which can be achieved to high accuracy by using separate supports for the cavity and the laser system and/or using a non-mechanical device for rotating the polarization etc.), stress-induced birefringence can be separated from the Lorentz-violation signal. 

Systematic effects in the proposed experiment would also include a change of the optical power coupled to the cavity as a function of the polarization angle. This would cause periodic temperature changes which, in turn, affect the cavity resonance frequency. The isotropic part common to both birefringent modes of that would drop out, as $c/(nL)$ is eliminated, but some anisotropic influence (birefringence due to an anisotropic temperature distribution) may remain. It can be minimized by using low optical powers and a fast rotation rate for the polarization, so that the cavity remains at a virtually constant average temperature. Another way to minimize this would be the simultaneous interrogation of the two birefringent modes, but then the relative fluctuations of the power in them could lead to a similar systematic effect. 

Bulk absorption of the material sets a lower limit for the linewidth $\Gamma$ of a matter-filled cavity. Since many systematic effects are proportional to $\Gamma$, it is desirable to keep $\Gamma$ small. For cavities made out of pure (crystalline) sapphire, a linewidth of $200$\,kHz has been reported \cite{Bra01}; quartz glass used in telecommunications fibers has absorption coefficients of below 0.5\,dB/km at a wavelength of 1.6\,$\mu$m \cite{Hobbs}, so cavities having well below 100\,kHz linewidth are in principle possible. This linewidth is lower than the one of cavities used in present Michelson-Morley experiments	 \cite{MuellerMM}. 

Earth's rotation gives a characteristic time-dependence to the different Lorentz violation coefficients in the laboratory frame. This allows to set separate limits on them and to distinguish the proper signal from most systematics. The time-dependence has been treated for the photon sector in Ref. \cite{KosteleckyMewes} and for the electron sector in Refs. \cite{ResSME,H2SME}. The signal for the proposed experiment can be derived from the material given therein and in Sec. \ref{n}.

The ultimate limit on the accuracy of cavity experiments is the shot-noise limit \cite{Hobbs}. For a 1\,nW of optical power impinging a cavity with 100\,kHz linewidth, the shot-noise limit would correspond to a frequency stability of the order of 1\,Hz, or $5\times 10^{-15}$, averaged over one second. In practice, the shot noise limit cannot be reached in cavity experiments operating at time-scales above approximately one second, where temperature fluctuations, material creep, drift of the laser power, residual amplitude modulation etc. limit the accuracy. However, the shot-noise limit could be approached more closely in the proposed experiment because of the ability of rotating the polarization on a faster time-scale and because influences to the cavity length would be common-mode and would be virtually removed. Thus, the proposed experiment might have an advantage over conventional cavity experiments, even the turntable versions. 

\section{Summary and Outlook}

We have eliminated assumptions on the electron wave-function from the theory underlying cavity tests of Lorentz invariance for both photons and electrons, thus completing the theory. This theory allows one to calculate the geometry change of crystals associated with Lorentz violation in the electrons' equation of motion, which leads to a measurable shift in the resonance frequencies of cavity oscillators. The sensitivity for experiments using various cavity materials was calculated, and materials providing exceptionally high sensitivity were identified. 

We then considered the shift in the index of refraction $n$ connected to Lorentz violation, that was found to consist of a density term and a dispersion term. The density term can be calculated exactly within the above theory for the geometry change. The dispersion term can be estimated for realistic situations. The shift in $n$ may have a significant influence on the signal of Lorentz tests. Also, we find that Lorentz violation causes birefringence in materials that are initially isotropic. 

Based on this theory, we reanalyzed experiments that have already been performed and found limits on Lorentz violating coefficients for the electron and the photon on levels of $10^{-15}$, and below for some electron coefficients. For the reader's convenience, we repeat the best limits derived herein in Tab. \ref{bestlim}. These limits represent an order of magnitude improvement for some electron parameters \cite{ResSME}, and the first limits on others. 

The parameters $(\tilde\kappa_{e-})^{ZZ}$ and $c_{ZZ}$ have not been measured yet, but can be measured, for example, in experiments similar to the above ones but using active rotation of the setup on a turntable. This generates an additional time-dependence that allows to measure the additional parameters. Setups optimized for obtaining simultaneous limits on photon and electron coefficients in optical experiments are discussed in \cite{ResSME}. New cavity experiments on Earth and in Space \cite{space} are now performed or proposed by several groups. Sucessors of the experiments described in Refs. \cite{cavitytests,MuellerMM,Wolfsme} using turntables are planned (some might even already be performed) and are expected to improve the sensitivity by two orders of magnitude \cite{CPT04} for the electrodynamic effects. These experiments use cavities with a dissimilar influence of the electron parameters, namely a sapphire cavity with the mode parallel to the $c$ axis, a circular sapphire cavity oriented orthogonal to the $c$-axis, and a niobium cavity. Combining these experiments should allow bounding both photon and electron parameters with the same improvement in accuracy. 

This work points several ways for obtaining increased sensitivity in experiments that use vacuum and matter-filled cavities, including an appropriate choice of materials that provides large length change coefficients and/or a strong sensitivity of the index of refraction. For example, a measurement of the symmetric parity-odd electron coefficients $c_{(0J)}$ using a calcite cavity would provide bounds that are 600 times better than a comparable experiment using a sapphire cavity. 

We also propose a new experiment that exploits the birefringence of isotropic materials caused by Lorentz violation. Since the birefringence depends on the same photon coefficients as $c$, the experiment could measure the same photon coefficients as a conventional Michelson-Morley type experiment. However, the thermal and other fluctuations that limit the accuracy in conventional cavity experiments could be eliminated.

\begin{table}
\caption{Summary of the best limits on photon and electron parameters derived herein. The experimental data used is from H. M\"uller {\em et al.} \cite{MuellerMM} and P. Wolf {\em et al.} \cite {Wolfsme}, unless otherwise noted. \label{bestlim}}
\begin{tabular}{lr}\hline
coefficient & limit $/10^{-15}$ \\ \hline
$(\tilde\kappa_{e-})^{YZ}$ & $0.52\pm 2.52$ \\
$(\tilde\kappa_{e-})^{XZ}$ & $-4.0\pm 3.3$ \\
$(\tilde\kappa_{e-})^{XY}$ & $-1.7\pm 1.6$ \\
$(\tilde\kappa_{e-})^{XX}-(\tilde\kappa_{e-})^{YY}$ & $2.8\pm 3.3$ \\ 
$c_{(YZ)}$ & $0.21\pm 0.46$ \\
$c_{(XZ)}$ & $-0.16\pm 0.63$ \\
$c_{(XY)}$ & $0.76\pm 0.35$ \\
$c_{XX}-c_{YY}$ & $1.15\pm 0.64$ \\ 
$|c_{XX}+c_{YY}-2c_{ZZ}-0.25(\tilde\kappa_{e-})^{ZZ}|$ & $\lesssim 1000$ \footnote{experimental data from Brillet and Hall \cite {BrilletHall} and M\"uller {\em et al.} \cite{MuellerMM}.} \\ \hline
\end{tabular}
\end{table}

\acknowledgments
I would like to thank Steve Chu as well as Sheng-wey Chiow, Sven Herrmann, Claus L\"ammerzahl, and Achim Peters for support and numerous discussions. Financial support by the Alexander von Humboldt-foundation is gratefully acknowledged.

\appendix
\section{Strong dispersion near a resonance}\label{largedispersion}

Here, we explore the region of strong dispersion for electromagnetic radiation having frequencies $\omega$ very close to a resonance frequency $\Omega$ of the medium, to find whether the influence of Lorentz violation is strongly enhanced for some frequencies. We assume that the detuning $\omega-\Omega$ is still large compared to the linewidth $\Gamma'$ of the resonance, so that we may use Eq. (\ref{dispersion}) for the susceptibility. However, the normalized dispersion $\bar n$ may become large, so that we cannot assume $1/(1+\bar n)\approx 1$ any more. For this section, we write the susceptibility as
\be\label{strongmodel}
\chi=\frac{K}{\Omega^2-\omega^2}+\chi_0\,
\ee
where 
\be
K=\frac{e^2}{\epsilon_0}\frac{N}{M}f\,.
\ee
$\chi_0=$const. sumarizes the susceptibility due to the other modes, which is approximately constant for the frequency range of interest here. We can now derive the index of refraction, the normalized dispersion and the term $d/(1+\bar n)$ as outlined above. Using the abbreviations
\begin{eqnarray}
A(\omega)&=& (3-\chi_0)(\Omega^2-\omega^2)-K\,,\nonumber\\ B(\omega)&=&2K+(2\chi_0+3)(\Omega^2-\omega^2)
\end{eqnarray}
we obtain
\begin{eqnarray}\label{largedisp}
n^2&=&\frac BA\,,\quad \bar n=\frac{9K\omega^2}{AB}, \nonumber \\ \frac{d}{1+\bar n}&=& \frac{-9\Omega^2K}{AB+9K\omega^2}\sum_a \frac 1\Omega \frac{\partial \Omega}{\partial \pi_a}\pi_a\,.
\end{eqnarray}
The dependence of $n^2$ and $\bar n$ on $\omega$ can be divided into three regions:
\begin{enumerate}
\item \label{low} $\omega^2<\omega^2_1\equiv \Omega^2-K/(3-\chi_0)$: Here, $n^2$ as well as $\bar n$ are positive ; they both diverge towards $+\infty$ at the high end of this range.
\item \label{med} $\omega^2_1<\omega^2<\omega^2_2\equiv \Omega^2+2K/(2\chi_0+3)$: In this region, $n^2$ is negative. At the high end of this region, $n^2=0$ and $\bar n\rightarrow -\infty$.
\item \label{high} $\omega^2>\Omega^2+2K/(2\chi_0+3)$: Here, $n^2$ as well as $\bar n$ are again positive.
\end{enumerate}
Let us study the case that the denominator $AB+9K\omega^2=0$ in the last Eq. (\ref{largedisp}). This may be interesting, because then the Lorentz violation parameters in the dispersion term are multiplied by an infinitely large factor. In this case, $\bar n=-1$. From the first Eq. (\ref{largedisp}), we obtain $n^2=B^2/(-9K\omega^2)<0$ (because both $K$ and $\omega$ are positive), i.e., this situation lies in the frequency range \ref{med}. A negative squared index of refraction, however, means that the electromagnetic wave inside the medium is evanescent (i.e., exponentially decaying), so that cavity resonances do not exist at frequencies in this region. Thus, unfortunately, this is not a way to obtain increased sensitivity in cavity experiments.

There are other potentially interesting situations: The extrema of the last Eq. (\ref{largedisp}) are at $\omega=0$ (minimum) and $\omega=\omega_1$ (maximum). We have $A(\omega_1)=0$ and $n^2(\omega_1)\rightarrow +\infty$, i.e., electromagnetic waves are propagating and thus cavity resonances may exist. The maximum has a value of 
\be\label{highsensleft}
\left. \frac{d}{1+\bar n}\right|_{\omega_1}=-\frac{\Omega^2(3-\chi_0)}{\Omega^2(3-\chi_0)-K}\sum_a \frac 1\Omega \frac{\partial \Omega}{\partial \pi_a}\pi_a\,.
\ee  
With appropriate $\Omega, \chi_0,$ and $K$, this may be large or even diverge. However, at frequencies near to $\Omega$, increased bulk absorption due to the imaginary part of the index of refraction will lead to an increase in the cavity linewidth $\Gamma$. This is undesirable, since many systematic effects \cite{MuellerMM} are proportional to that linewidth. The increase of $\Gamma$ relative to its value $\Gamma_0$ without bulk absorption is given by \cite{Bra01}
\be
\Gamma(\omega)=\Gamma_0+\frac{18n^2(\omega)}{(n^2(\omega)+2)^2}\Gamma'\,.
\ee
This equation holds as long as the detuning from resonance is much larger than the linewidth of the material resonance, $|\omega-\Omega|\gg \Gamma'$. In particular, we obtain 
\be
\Gamma(\omega_1)=\Gamma_0+\frac 2K(3-\chi_0)[(3-\chi_0)\Omega^2-K]\Gamma'\,.
\ee
Comparing to Eq. (\ref{highsensleft}) shows that in principle the parameters can be chosen such that the sensitivity becomes high while the increase in linewidth is low. However, for a strong enhancement of the sensitivity, what is necessary is a material whose $n$ approaches infinity while the absorption remains low, and where $n$ diverges in such a way that Eq. \ref{highsensleft} becomes large at the same time. It is a challenge to find a practical material which realizes this ``fine-tuning" of the dispersion curve. Even if such a material would be found, the formal gain in sensitivity would have to be weighed against issues like the temperature coefficient of $n$ (that is likely to diverge together with $n$) to see whether such an experiment is worthwile.

Especially strong dispersion is seen in ``slow light" experiments \cite{Hau99}. Here, $n\approx 1$ and $\bar n\gg 1$, which corresponds to a low group velocity $v_g=c/[n(1+\bar n)]$. It immediately follows from Eq. (\ref{totaleffect}), however, that a large $\bar n$ leads to suppressed, rather than enhanced, sensitivity in the cavity experiments described herein.

\section{Lorentz-violating birefringence in isotropic crystals}
\label{LVB}

We consider a cubic ionic crystal of NaCl structure that has a lattice constant $a$. The primitive direct lattice vectors $\vec l_k$ ($k=1,2,3$) are contained in the matrix $l_{jk}=a\delta_{jk}$; the equilibrium positions of the ions can be described by a linear combination $n_k l_{jk}$. For calculating the index of refraction $n$, we need to calculate the resonance frequencies $\Omega$ of elastic waves. As explained above, a Lorentz-violating shift in $\Omega$ will be given by a modification of the potential energy for small dilations of the ions $\vec r\rightarrow \vec r+\Delta \vec r$ from their equilibrium positions $\vec r$. This corresponds to dilations $l_{jk}\rightarrow l_{jk}+\tilde l_{jk}$) of the primitive lattice vectors. The Hamiltonian can be be written as
\be 
H=\sum_{e-}\frac{p_e^2}{2m}+\sum_{I}\frac{p_I^2}{2\bar M}+\Phi\,,    
\ee
where the first term is the electrons' kinetic energy $T_e$, summed over all the electrons, the second the kinetic energy of the kinetic energy of the atom cores, and $\Phi$ the Coulomb potential, summed over all charges. 

Lorentz violation in the electron's equation of motion replaces the electron's kinetic energy by $(\delta_{ij}+2E'_{ij})p_ip_j/(2m)$. We can assume that the electronic states are changed adiabatically by the lattice vibration, i.e., their kinetic energy effectuates an effective potential energy as a function of the dilations of the atomic cores. The electron wave functions can be given by Bloch functions; the expectation value
\be
\sum_e \frac{p_e^2}{2m}=\frac{\hbar^2N_{e,u}}{2m}(\delta_{ij}+2E'_{ij})\oxi_{lk} k_{il}k_{jk}
\ee
for the kinetic energy of the electrons within a single unit cell (containing $N_{e,u}$ electrons) has been calculated in Ref. \cite{ResSME}. From Eqs. (\ref{kappadef},\ref{kappalambda}), we obtain
\be
\oxi_{lk}=\frac{m|\det(l_{ij})|}{4\pi^2N_{e,u}\hbar^2}l_{ak}l_{ql}\lambda_{ddqa}\,.
\ee
For a cubic crystal with a lattice constant $a$, we have the simplified relations
\be
l_{ij}=a\delta_{ij}\,,\quad k_{ij}=\frac{2\pi}{a}\delta_{ij}\,,\quad \oxi_{lk}=\frac{ma^5S_1}{4\pi^2N_{e,u}\hbar^2}\delta_{lk}\,,
\ee
where $S_1=S_{11}+S_{12}+S_{13}$ is given by the elastic components, and $a^3=|\Det(l_{ij})|$ expresses the volume of the unit cell. The dilations $\tilde l_{ij}$ of the direct lattice vectors will lead to small dilations of the reciprocal lattice. In the special case of cubic crystals it is given by
\be
k_{ij}=\frac{2\pi}{a}\left(\delta_{ij}-\frac 1a \tilde l_{ji}\right)\,.
\ee 
Calculating the expectation value of the electrons' kinetic energy gives
\be
T_e=\frac 12a^3S_1(\delta_{ij}+2E'_{ij})\left(\delta_{ij}-\frac 2a\tilde l_{ji}+\frac{1}{a^2}\tilde l_{ki}\tilde l_{jk}\right)
\ee
or 
\begin{eqnarray}\label{Te}
T_e&=&\frac 12a^3S_1\left(3+2\tr(E')-(2+4\tr(E'))\frac{\Delta V}{V} \right. \nonumber \\ & & \left. +\frac{1}{a^2} [(\Delta a)^2+ 2 \Delta \vec r E' \Delta \vec a ]\right)\,,
\end{eqnarray}
where we used
\be
\tilde l_{ii}=a \frac{\Delta V}{V} \,,\quad \tilde l_{li}\tilde l_{li}=(\Delta a)^2\,.
\ee

For the Lorentz-violating Coulomb contribution per unit cell, we write Eq. (\ref{coulombsme}) as
\be
\Phi=\frac 12 \frac{Z^2e^2}{4\pi}\sum (\pm)\left[\frac 1r+\frac{r_i r_j (\kappa_{DE})_{ij}}{2r^3}\right]
\ee
where the sum is running over all other ions contained in the lattice ; $r$ is the distance of the ions relative to each other. The symbol $(\pm)$ indicates the positive and negative charges of the ions. The factor of $1/2$ corrects for the double-counting of the ions in the summation. 

The distances $\vec r$ can be written as a linear combination of the lattice vectors. In NaCl structure, the distance of next neighbors is diagonal in the unit cell, with a length $R=a/\sqrt{2}$. For the following calculations, we use cartesian coordinates parallel to these distances and assume a dilation $\epsilon$ along, e.g., the $z$-direction. To write $\Phi$ explicitly as a function of the dilations, we expand it to second order in $\epsilon$:
\begin{eqnarray}
\Phi&=&\frac{Z^2e^2}{8\pi}\sum(\pm)\left[\frac 1r-\frac{z\epsilon}{r^3}-\frac{\epsilon^2}{2r^3}+\frac 32 \frac{z^2\epsilon^2}{r^5} \right. \nonumber \\ & & \left. +\frac12\left(r^2\tr(\kappa_{DE})+2\epsilon z (\tilde \kappa_{DE})_{zz}+\epsilon^2(\tilde\kappa_{DE})_{zz}\right) \right. \nonumber \\ && \left. \times \left(\frac{1}{r^3}-3\frac{z\epsilon}{r^5}-\frac 32 \frac{\epsilon^2}{r^5}+\frac{15}{2}\frac{z^2\epsilon^2}{r^7}\right) \right]+\mathcal O(\tilde \epsilon^3)\,.\quad\quad
\end{eqnarray}
We consider elastic waves with a wavelength much longer than the lattice constant $a$. Thus, $\Delta\vec r$ can be regarded as fixed for all ions and taken out of the summation. In the summation, all terms that are proportional to even powers of $z$ drop out. For the other terms, we define the dimensionless quantities \cite{sumexplicit}
\begin{eqnarray}
\alpha_1 &\equiv& R \sum\frac{(\pm)}{r}\approx -1.74 \,,\quad  \alpha_3\equiv R^3\sum\frac{(\pm)}{r^3}\approx -3.24 \,, \nonumber \\ \alpha^2_5 &\equiv& R^3\sum(\pm)\frac{r_1^2}{r^5}\approx-1.08
\end{eqnarray}
(the numerical values suggest $3\alpha^2_5=\alpha_3$, at least to the accuracy required here.) This allows to carry out the summation over all ions:
\begin{eqnarray}
\Phi& = & \frac{Z^2e^2}{8\pi R}\left[\alpha_1\left(1+\frac 12 \tr(\tilde\kappa_{DE})\right) \right. \nonumber \\ & & \left. +\frac{\alpha_3\epsilon^2}{2R^2}\left(-1+(\tilde\kappa_{DE})_{zz}-\frac 32 \tr(\tilde\kappa_{DE})\right) \right. \nonumber \\ & & \left. +\frac{3\alpha^2_5\epsilon^2}{R^2}\left(\frac 12-(\tilde\kappa_{DE})_{zz}+\frac{5}{4}\tr(\tilde\kappa_{DE})\right)\right]\,.\quad 
\end{eqnarray}
For comparing the contributions of $T_e$ and $\Phi$, note that $a=\sqrt{2}R$, and $\delta a=\sqrt{2} \epsilon$. For the terms that are quadratic in $\epsilon$, we thus obtain 
\begin{eqnarray}
T_e+\Phi& = & \sqrt{2}RS_1\epsilon^2(1+2E'_{zz})+\frac{Z^2e^2}{8\pi R^3}\epsilon^2 \left[-\frac{\alpha_3}{2}+\frac 32\alpha^2_5 \right. \nonumber \\ & &  \left.+\left(\frac{\alpha_3}{2}-3\alpha^2_5\right)(\tilde\kappa_{DE})_{zz}\right]\,.
\end{eqnarray}
We left out the term proportional to $\tr(E')$, since it is not of interest for usual experiments due to its rotation-invariance. We can write the two factors of dimension energy as
\be
-\eta \frac{Z^2e^2\alpha_1}{8\pi R}=\sqrt{2}R^3S_1\,,
\ee
where $\eta$ is a dimensionless factor. From the data on the 16 materials given in Tab. 7 of Ref. \cite{Kittel}, we find $2.4\leq \eta \leq 4.2$, with an average of $\eta \approx 3.5$. We can thus write (using $\alpha_3=3\alpha^2_5$) 
\be
T_e+\Phi= \sqrt{2}RS_1\epsilon^2\left[1+2E'_{zz}+\frac{\alpha_3}{2\alpha_1\eta}(\tilde\kappa_{DE})_{zz})\right].
\ee

The frequency $\Omega=\sqrt{k/\bar M}$ of the corresponding resonance will be given by square root of the coefficient $k$ in front of $\epsilon^2$, divided by the effective mass $\bar M$ of the mode. Hence, the Lorentz violation leads to a change
\be
\frac{\delta \Omega}{\Omega}=\frac 12(\tilde\kappa_E)_{zz}
\ee
in the resonance frequency, where
\be
(\tilde\kappa_E)_{zz}=2E'_{zz}+\frac{3\alpha_3}{2\alpha_1\eta}(\tilde\kappa_{DE})_{zz}\,.
\ee 
As above, this leads to a shift in the index of refraction of $x$-polarized light; for light propagating into the $z$-direction, for example there will be a difference 
\be
\frac{\delta n}{n} =\frac \beta 2 [(\tilde\kappa_E)_{11}-(\tilde\kappa_E)_{22}]
\ee 
between the index of refraction for $x$ and $y$ polarized light. For $E'=0$, i.e., Lorentz violation in the Maxwell sector only, we recover Eq. (\ref{omipchange}) with the numerical factor in front of $\tilde\kappa_{DE}$ being $3\alpha_3/(4\alpha_1\eta)\approx 0.40$ instead of $1/4$. This is due the combined influence of $T_e$ and $\Phi$ and the geometry-dependent factors $\alpha_1$ and $\alpha_3$.


\begin{references}
\bibitem{CMremark} The Clausius-Mosotti equation describes the polarization of any one spherical region within the solid due to both the external electric field and the ``internal" field caused by the polarization of the surrounding ions. It works best for materials where the averaging over a sphere gives a good estimate of the internal field, i.e., isotropic materials.
\bibitem{sumexplicit} E.g., explicitly
\be
\alpha^2_5=\sum_{l=-\infty}^\infty\sum_{m=-\infty}^\infty\sum_{n=-\infty}^\infty\frac{(-1)^{l+m+n}n^2}{(l^2+m^2+m^2)^{5/2}}\,,
\ee
where $l=m=n=0$ is excluded from the summation.
\end{references}
\end{document}